\documentclass[fleqn,11pt]{SelfArx} 

\usepackage[english]{babel} 

\usepackage{lipsum}

\definecolor{color1}{RGB}{0,0,0} 
\definecolor{color2}{RGB}{20,100,100} 

\usepackage{hyperref} 

\usepackage{graphicx,caption}

\newsavebox{\arrangebox}

\usepackage{textcomp}
\usepackage{ragged2e}

\usepackage{cases}

\usepackage[backend=bibtex,bibencoding=ascii,style=phys,articletitle=true,biblabel=brackets,giveninits=true, sorting=none]{biblatex}
\usepackage{color}

\usepackage{subfig}
\usepackage{amsmath}

\usepackage{stfloats}

\usepackage{extdash}

\JournalInfo{Preprint Paper Drafts, IAP Jena \the\year{}}
\Archive{Johannes Kaufmann}

\PaperTitle{Ultra-shallow EUV and soft X-ray gratings fabricated by broad-beam nitrogen ion irradiation
}

\Authors{Johannes Kaufmann\textsuperscript{1}, Thomas Siefke\textsuperscript{1,2,3}, Uwe Zeitner\textsuperscript{1,2}\\ \scriptsize{\textsuperscript{1}Friedrich Schiller University, Institute of Applied Physics, Albert-Einstein-St. 15, 07745 Jena, Germany\\
\textsuperscript{2}Fraunhofer Institute for Applied Optics and Precision Engineering, Albert-Einstein-St. 7, 07745 Jena, Germany\\
\textsuperscript{3} Ernst-Abbe-Hochschule Jena University of Applied Sciences, Carl-Zeiss-Promenade~2, 07745 Jena, Germany\\
}
}

\newcommand{\keywordname}{Keywords}

\Abstract{{Controlled and precise fabrication of structures with heights in the range of single digit nanometres is one of the challenges for diffraction gratings operating near-normal incidence in the extreme ultraviolet (EUV) and soft X-ray range.
		Here, we expand on previous research utilizing swelling of silicon after irradiation with ions as alternative to conventional dry etching.
		By irradiating silicon through a mask with a broad beam of nitrogen ions, we realized lamellar gratings in a precise and well controlled process.
		We were able to fabricate gratings with structure heights between 1.00\,$\pm$\,0.05 to 10\,$\pm$\,0.5\,nm and a pitch of 1\,{\fontfamily{ppl}\selectfont µ}m, which is suitable for both EUV and soft X-ray applications.
		A variation of ion energy from 20 to 40\,keV further expands the foundations of this process and yielded an additional parameter to control the resulting structure height and shape.
	}
}

\begin{document}
\setlength{\abovedisplayskip}{0pt}

\maketitle

\thispagestyle{empty}

\vspace{2pc}

\maketitle

\section{Introduction}
Prominently, light from the EUV and X\Hyphdash*ray range is an important tool for lithography\textsuperscript{\tiny{\cite{Levinson2022,Mojarad2015,Fu2023,Wu2023}}}
as well as imaging and metrology applications, which utilize the wavelength range for fluorescence\textsuperscript{\tiny{\cite{Hoenicke2023,Ciesielski2023}}}, scattering\textsuperscript{\tiny{\cite{Ruben2022,Ciesielski2023,Chuang2022,Weinhardt2024}}}, spectroscopy\textsuperscript{\tiny{\cite{Chuang2022,Weinhardt2024,Shatokhin2018,Holburg2025}}}, pump-probe experiments for ultrafast dynamics\textsuperscript{\tiny{\cite{Jiang2013,Wang2023}}} and other applications\textsuperscript{\tiny{\cite{Eschen2022,Chen2022,Aidukas2024,Battistelli2024,Gutberlet2023,Grote2022}}}.
In these applications a large variety of different gratings is utilized\textsuperscript{\tiny{\cite{Mojarad2015,Poletto2018,Ruben2022,Wu2023,Fu2023,Ciesielski2023,Gutberlet2023,Li2024,Grote2022,Chuang2022,Weinhardt2024,Shatokhin2018,Holburg2025}}}.
However, in this spectral range operation is often in grazing incidence, which demands long optical paths and requires large device footprints for expensive and big optical elements\textsuperscript{\tiny{\cite{Poletto2018,Wu2023,Li2024}}}.
With respect to application in compact tabletop devices an operation at near normal incidence would thus be highly advantageous\textsuperscript{\tiny{\cite{Eschen2022,Holburg2025}}}.
In the desired spectral range around 1\,nm to 50\,nm this requires techniques to precisely produce well controlled structure heights in the range of a few nanometres for ultra shallow gratings.
Especially the controllability of this parameter is non\Hyphdash*trivial, as conventional dry etching methods remove material
at tens of nanometres per minute -- much larger than the desired elements -- while increasing the roughness of the substrate\textsuperscript{\tiny{\cite{Williams2003,Wu2010,Dowling2016,Huff2021,RackaSzmidt2022,Miles2022}}}.
Additionally, an operation at near normal incidence commonly requires gratings to be coated with a reflective multilayer.
Designing such a multilayer to tolerate large uncertainties in the structure height greatly restricts the available degrees of freedom, ultimately resulting in lower performance.
In this context, simulations highlight the demand for alternative processes achieving precisely controlled structure heights\textsuperscript{\tiny{\cite{Yang2017a,Koike2023,Koike2024a}}}.
This demand is addressed by the technique for the fabrication of ultra shallow optical elements presented here.

Our approach relies on
the fabrication of surface structures and gratings utilizing swelling and sputtering during irradiation with ions for varying ion--substrate combinations\textsuperscript{\tiny{\cite{Kaufmann2024,Garcia2022,Huang2022,Miles2022,Chen2021,Li2021,Jensen2008,Lindner2009,Momota2019,Zhou2019,Kaufmann2025}}}.
Previous research has already established that due to processing times, areas beyond {\fontfamily{ppl}\selectfont µ}m\textsuperscript{2} cannot be structured utilizing a direct write process with a focused ion beam\textsuperscript{\tiny{\cite{Huang2022, Garcia2022}}}.
Consequently, feasible processing times require alternative methods capable to utilize broad ion beams.
Examples of this include self organization for specific ion--substrate combinations\textsuperscript{\tiny{\cite{Chen2021, Li2021}}} or the utilization of orientation dependent sputter rates\textsuperscript{\tiny{\cite{Miles2022}}}.
But currently, both of these approaches are limited in the usable substrates and the fabricated structures often lack uniformity in combination with increased roughness.
To circumvent this, irradiation through stencil masks can be utilized to achieve a selective shadowing of the substrate\textsuperscript{\tiny{\cite{Jensen2008,Lindner2009,Momota2019,Zhou2019}}}.
However, these methods had limited lateral feature sizes of above 10\,{\fontfamily{ppl}\selectfont µ}m for lamellar gratings with transition regions between masked and irradiated areas of around 1\,{\fontfamily{ppl}\selectfont µ}m, setting a limit to possible structure shapes.

To enhance the available selection of fabrication techniques, we present further insights into a previously established
efficient process for grating fabrication relying on a broad beam of ions and a photoresist mask on the substrate\textsuperscript{\tiny{\cite{Kaufmann2025,Kaufmann2024,Kaufmann2025a,Kaufmann2025b}}}.
Consistent with our previous experience, fabricated structures were characterized via atomic force microscopy (AFM).
Here, we utilize nitrogen ions on a silicon substrate due to its common use, widespread availability and well known amorphisation behaviour of silicon in conjunction with the reduced lateral spread of nitrogen ions in comparison to lighter ions\textsuperscript{\tiny{\cite{Pelaz2004,Ziegler2012}}}.
This alternative approach to area\Hyphdash*selective ion irradiation effects of amorphisation and volume change caused by ion irradiation enabled us to efficiently fabricate lamellar gratings with small transition region and unprecedented control over the resulting structure height.

\section{Experimental Procedure}
Presented samples were irradiated using our 4 grid accelerator broad ion source (4GABIS).
The linear accelerator 4GABIS achieves acceleration voltages of up to 40\,keV.
Here, this was utilized to accelerate single charged nitrogen molecules (N\textsubscript{2}\textsuperscript{+}) from a plasma source with an energy between 20~to~40\,keV.
With the assumption that N\textsubscript{2}\textsuperscript{+} ions split into parts of equal energy upon hitting a target this equates to an irradiation with nitrogen with an energy of 10~to~20\,keV\textsuperscript{\tiny{\cite{Jacob1998}}}.
Unless specified, the following discussion will reference experiments with the energy of the N\textsubscript{2}\textsuperscript{+} ions.
Depending on energy, samples were irradiated with ion current densities between 45~and~75\,{\fontfamily{ppl}\selectfont µ}A/cm\textsuperscript{2}, increasing from 20~to~40\,keV, respectively.
Adjusting irradiation time between 30~and~500\,s resulted in a variation of fluence in the range of~2.4\,$\cdot$\,10\textsuperscript{16} to
3.7\,$\cdot$\,10\textsuperscript{17}\,ions\,$\cdot$\,cm\textsuperscript{-2}.
Samples were connected to the water cooled target stage via backside cooling with nitrogen to improve heat conduction, limiting the maximum observed temperature of the samples to below 85\,$^\circ$C.
During experiments, samples were heated to 70\,$^\circ$C within the span of a minute, followed by an approximately linear increase by 2\,K\,per\,minute.

As substrates the experiments used superpolished Si(100)-wafers with an initial surface roughness of around 0.3\,\AA~provided by \textsc{Siltronic AG}.
During ion irradiation a polymer with a binary lamellar structure was utilized to mask parts of the substrate.
To
gain insight into the fundamental characteristics of the process we focused on a pitch of 1\,{\fontfamily{ppl}\selectfont µ}m.
\begin{figure}[b]\centering
	\includegraphics[width=0.45\textwidth]{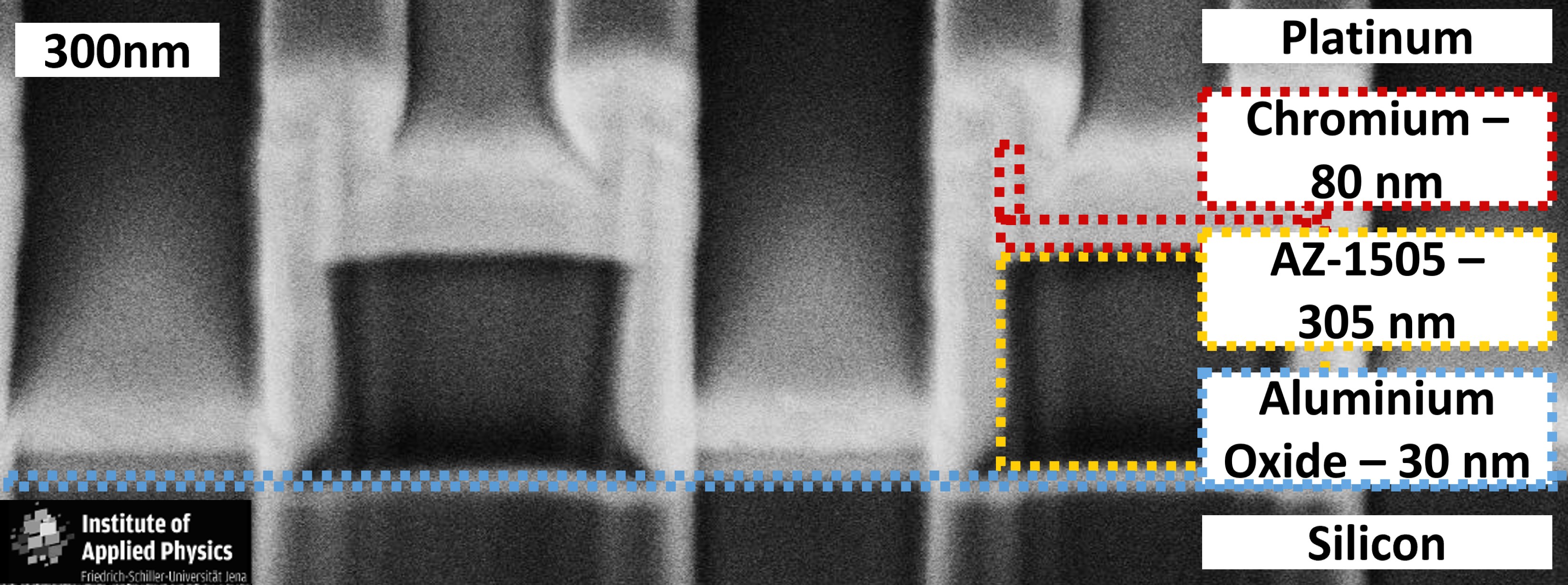}
	\caption{\small{SEM image of a cross section of the utilized mask. The platinum on top originates from the sample preparation for the cutting process.}}
	\label{fig:Mask}
\end{figure}
\noindent
Figure~\ref{fig:Mask} shows a scanning electron microscope (SEM) cross section of the mask prior to irradiation.
From the silicon substrate at the bottom to the top it consists of approximately 30\,nm aluminium oxide acting as protection layer, a 305\,nm AZ-1505 photoresist mask and approximately 80\,nm chromium from the fabrication process.
The fabrication process of this mask is based on electron beam lithography and has been discussed in more detail previously\textsuperscript{\tiny{\cite{Haedrich2022,Kaufmann2025}}}.
In comparison to the previous experiments, here we reduced the thickness of the AZ-1505 photoresist of the mask
to 305\,nm, as the heavier nitrogen ions have a lower range within the mask than the previously utilized helium ions.
This reduces the aspect ratio and fabrication time while still preventing any ions from reaching the substrate via penetration through the mask.
The finalized sample has a patterned area of 5\,mm by 5\,mm in the centre of 15\,mm by 15\,mm silicon chips completely masked by resist and chromium outside of the grating area.

As the irradiation of silicon with nitrogen can lead to the formation of Si\textsubscript{x}N\textsubscript{y}\Hyphdash*compounds the mask removal process described in\textsuperscript{\tiny{\cite{Kaufmann2025}}} was adapted from previous experiments to minimize the exposure to phosphoric acid, which is known to etch silicon nitride\textsuperscript{\tiny{\cite{Williams2003}}}.
For this purpose, first the chromium was stripped
in a wet chemical process, followed by the removal of the resist layer in an oxygen plasma.
Afterwards, the protective aluminium oxide layer is removed by treatment with phosphoric acid at 50\,$^\circ$C for 3\,minutes, thus minimizing potential etching.
As a final step before characterization, samples were cleaned utilizing Caro's acid, ammonia and deionized water in conjunction with megasonic.

For the characterization a \textsc{Dimension Edge}\textsuperscript{\tiny\texttrademark} AFM from \textsc{Bruker Corporation} was used.
Our measurements were performed in tapping mode at ambient conditions with a Tap\Hyphdash*300~Al\Hyphdash*G~tip purchased from \textsc{BudgetSensors} as AFM probe.
This AFM probe has a tip radius of approximately 10\,nm and a force constant of 40\,N/m.
The measurements were conducted on 10\,{\fontfamily{ppl}\selectfont µ}m by 10\,{\fontfamily{ppl}\selectfont µ}m areas with a resolution of 10\,nm per pixel in the fast direction.
The high force constant of the tip allowed us to reduce the uncertainty of the height measurement at the AFM to the limits of controller noise experienced by the tip, which resulted in a total measurement uncertainty of around 30\,pm for the height measurement\textsuperscript{\tiny{\cite{Butt1995}}}.
All measurements of roughness reported here refer to the root mean square (Sq) and include correction for the controller noise.

\section{Results and Discussion}
Utilizing the swelling of silicon caused by irradiation with nitrogen ions we were able to realize shallow gratings with a
lamellar structure.
\begin{figure*}[b]\centering
	\null
	\subfloat[]{\begin{minipage}{0.5\textwidth}
			\null\hfill
			\includegraphics[width=0.9\textwidth]{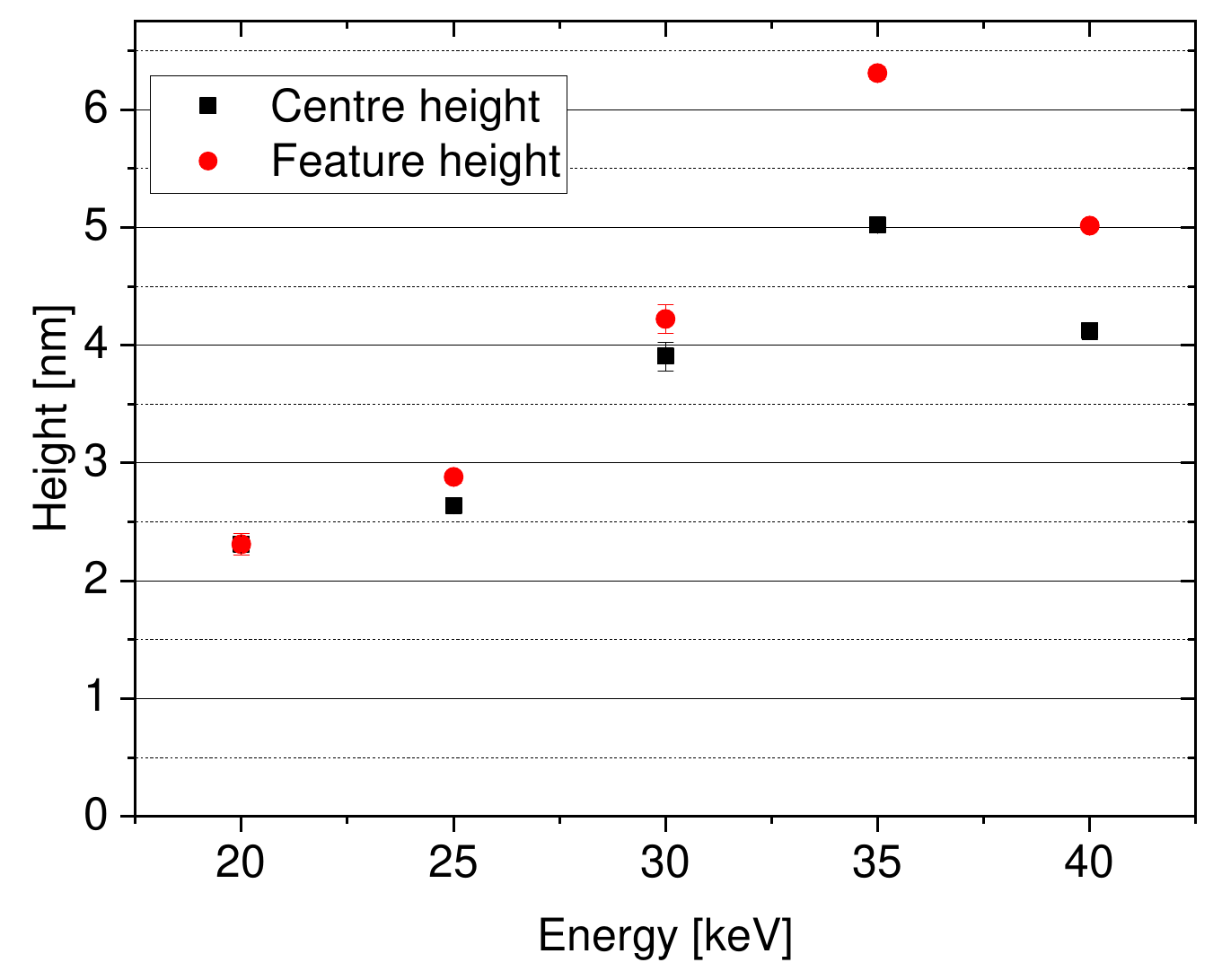}
			\hfill\null
	\end{minipage}}
	\subfloat[]{\begin{minipage}{0.5\textwidth}
			\null\hfill
			\includegraphics[width=0.97\textwidth]{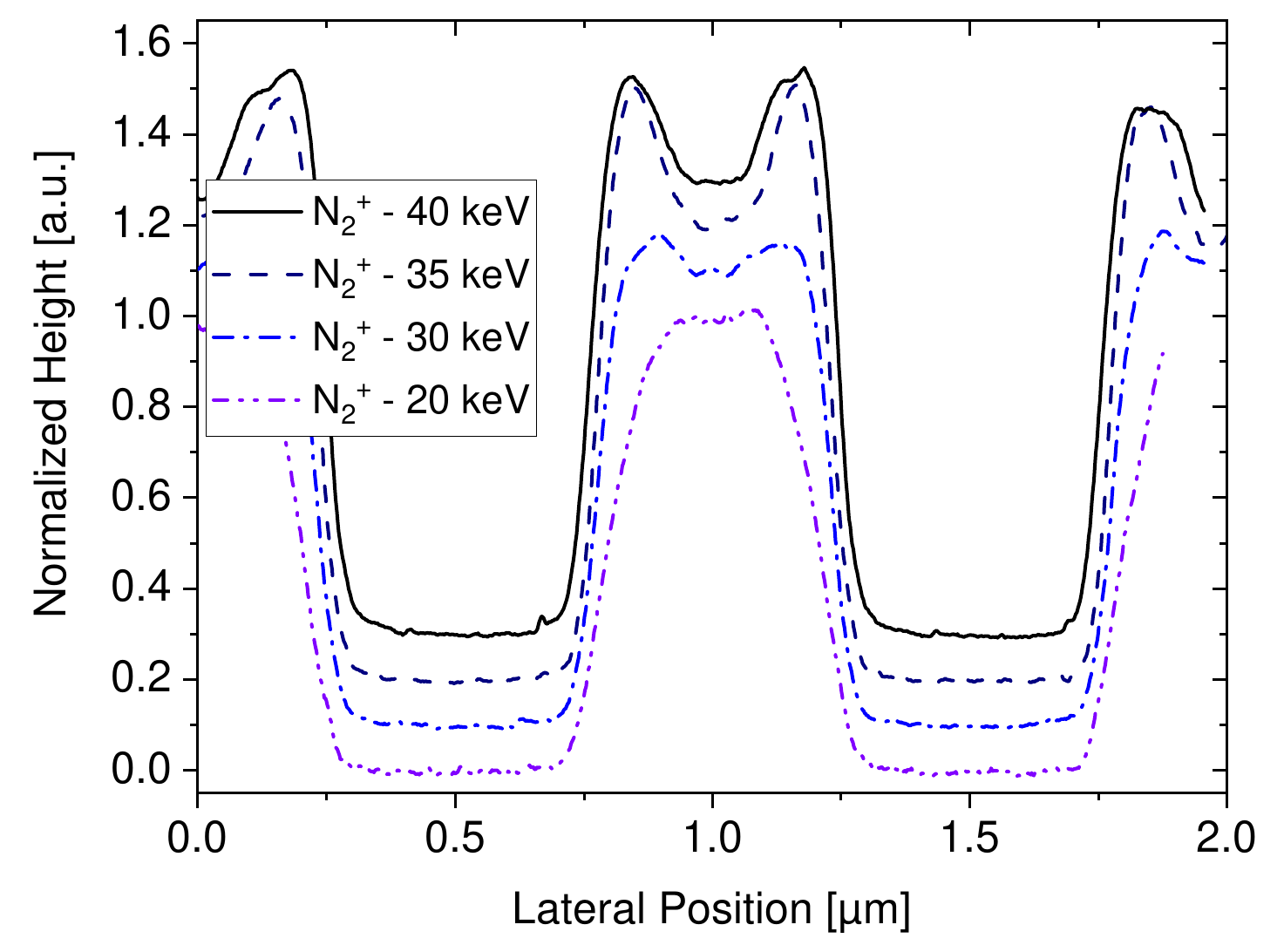}
			\hfill\null
	\end{minipage}}
	\null

	\caption{\small{Change in \textbf{(a)} structure height for irradiations with similar fluence and \textbf{(b)} structure shape in dependence of acceleration energy.
			The normalization sets the centre of the structure to a height of 1. Curves are vertically displaced for better visibility.}}
	\label{fig:Energy-var-1}
\end{figure*}
\noindent
\begin{figure*}[tb]\centering
	\subfloat[]{
	\includegraphics[height=2.98cm]{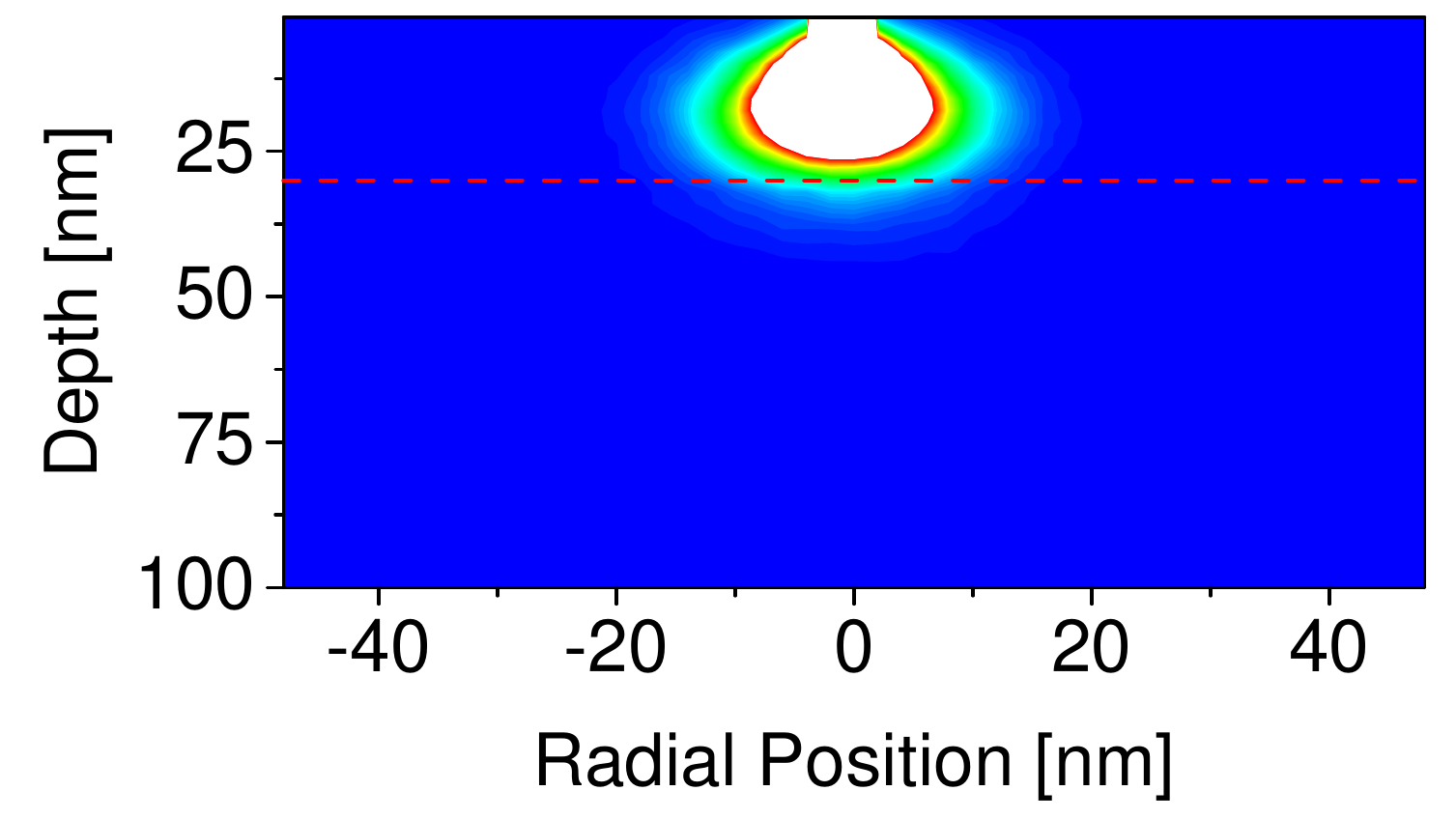}
}
\subfloat[]{
	\includegraphics[height=2.98cm]{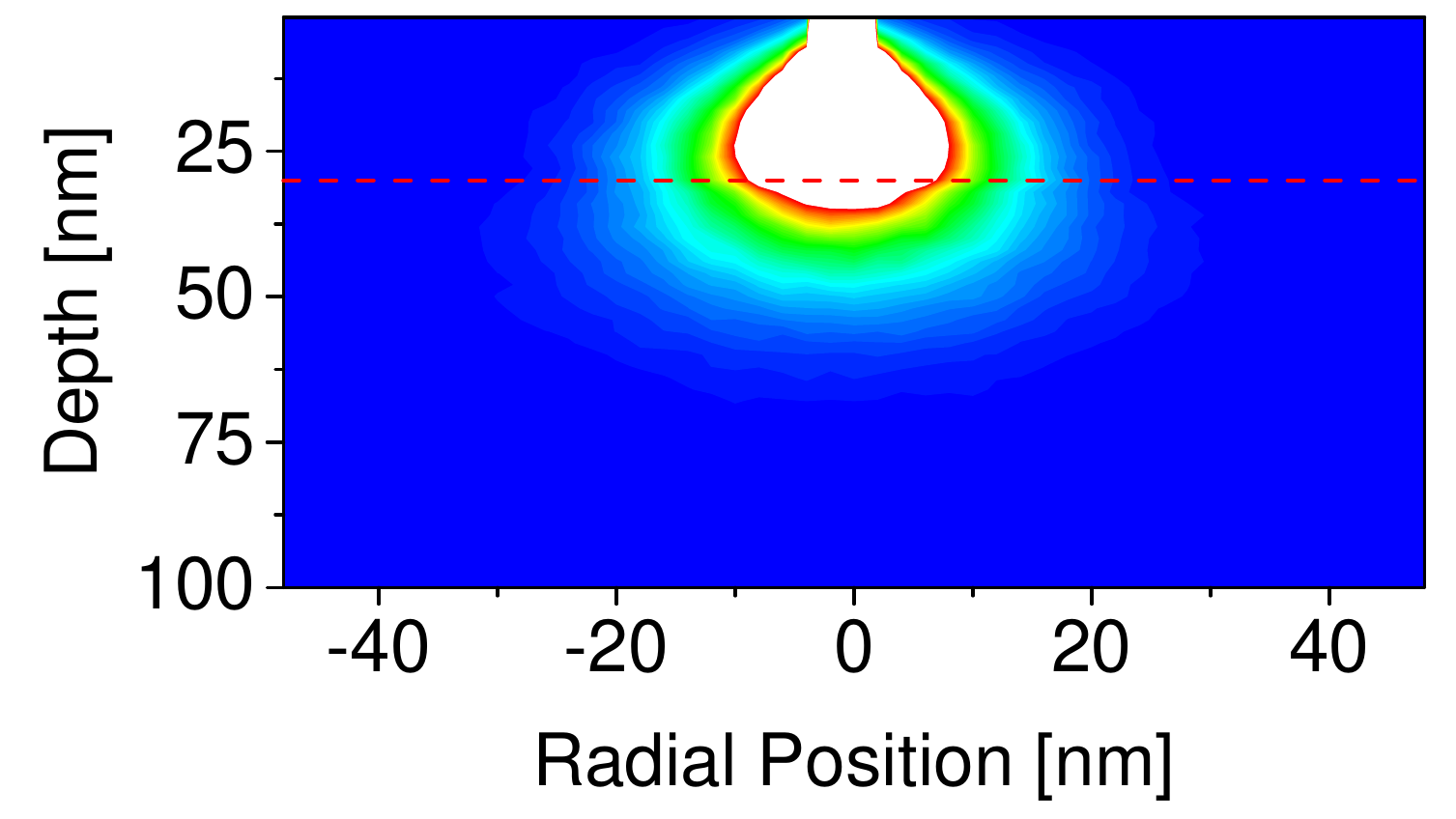}
}
\subfloat[]{
	\includegraphics[height=2.98cm]{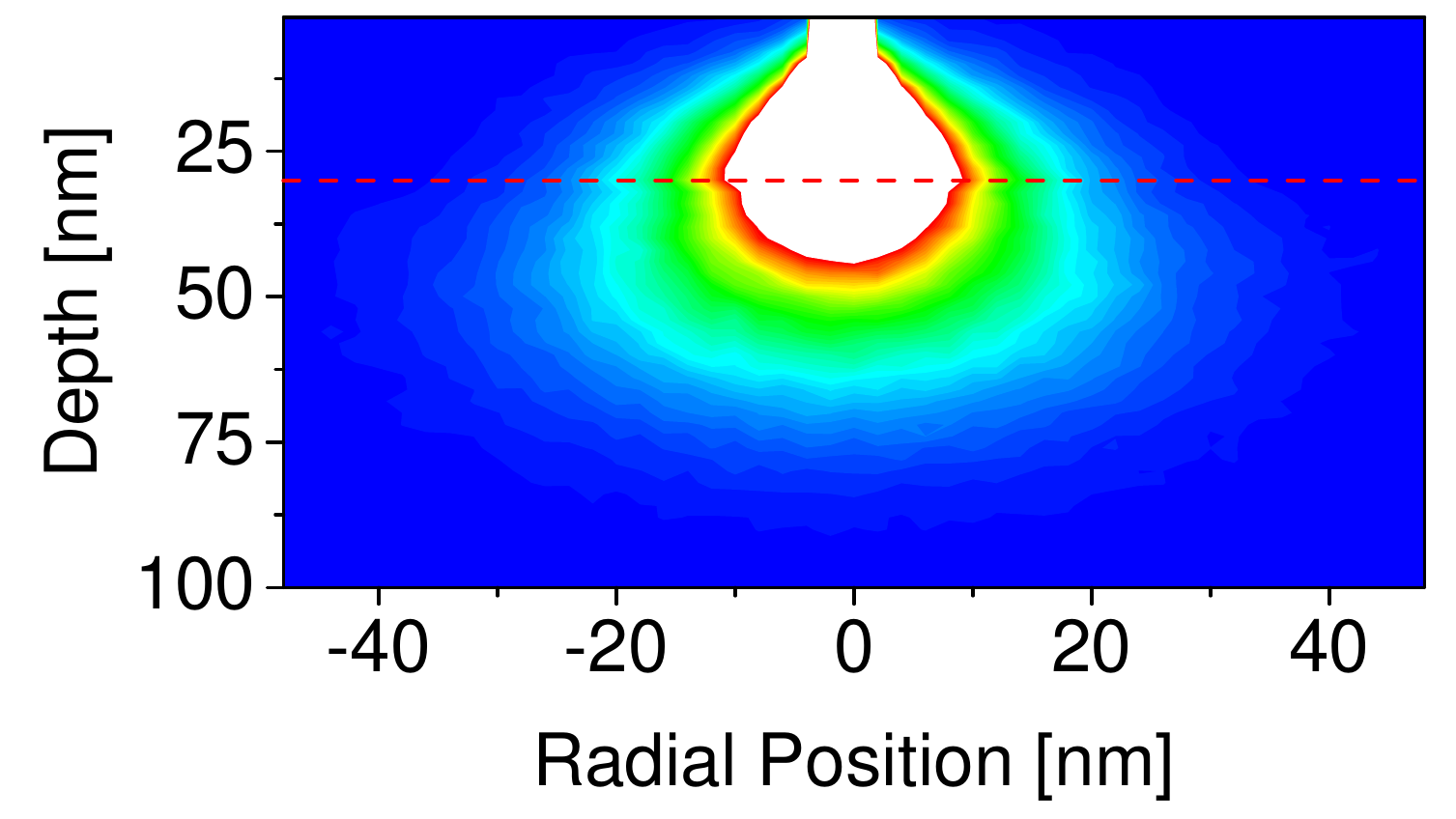}
}
\subfloat[]{\raisebox{0mm}{
	\includegraphics[height=2.98cm]{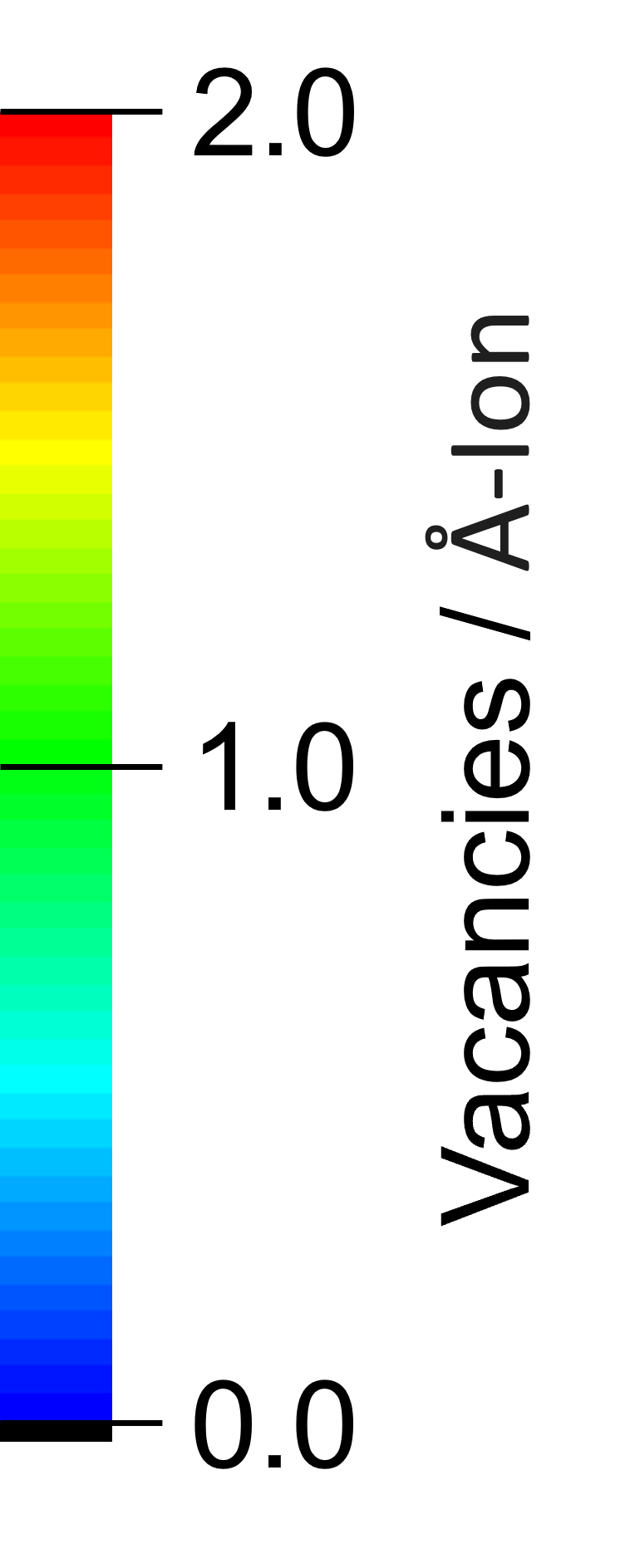}
	}
}
\null
	
	\null
	\subfloat[]{
		\includegraphics[height=2.98cm]{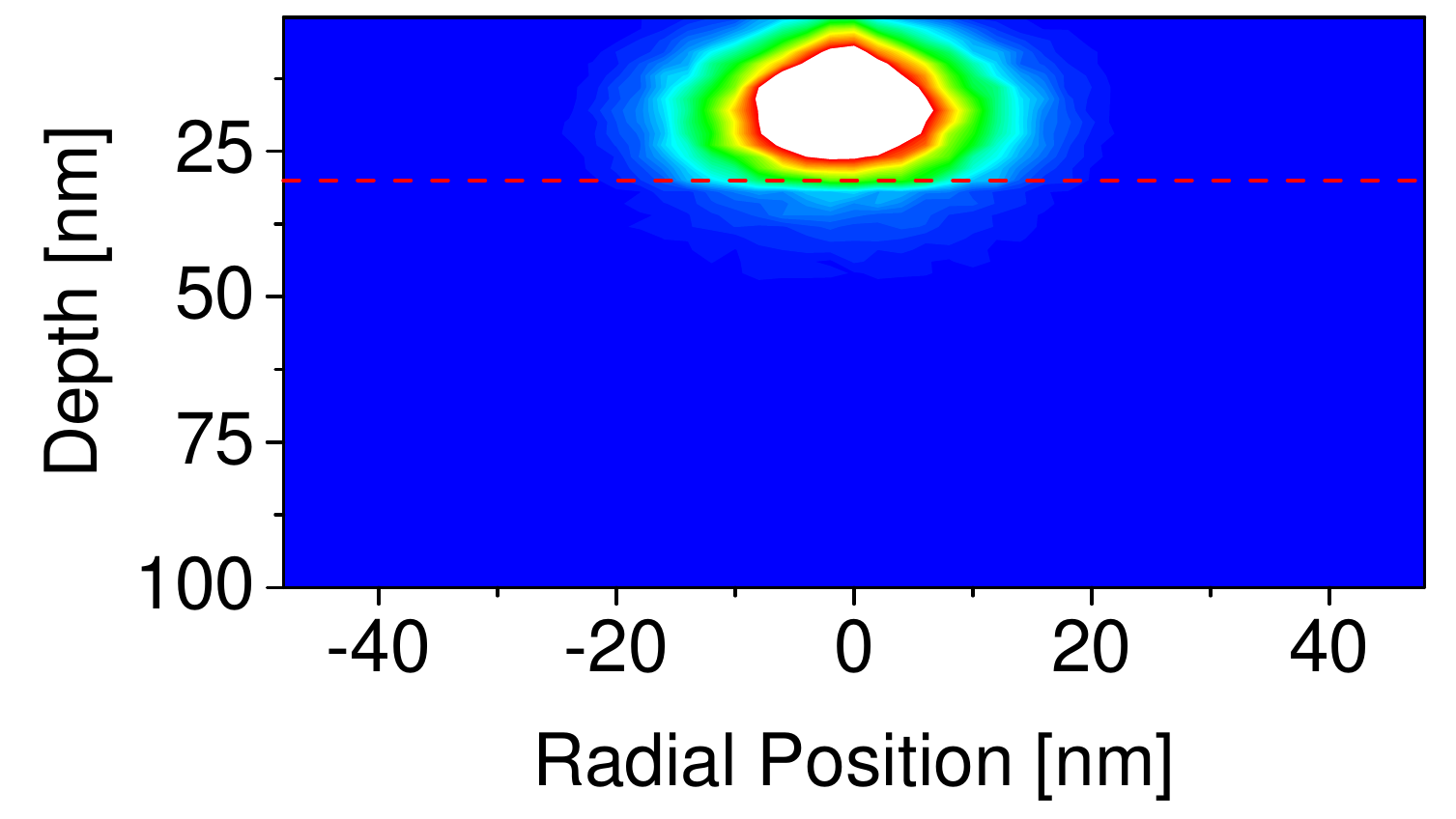}
	}
	\subfloat[]{
		\includegraphics[height=2.98cm]{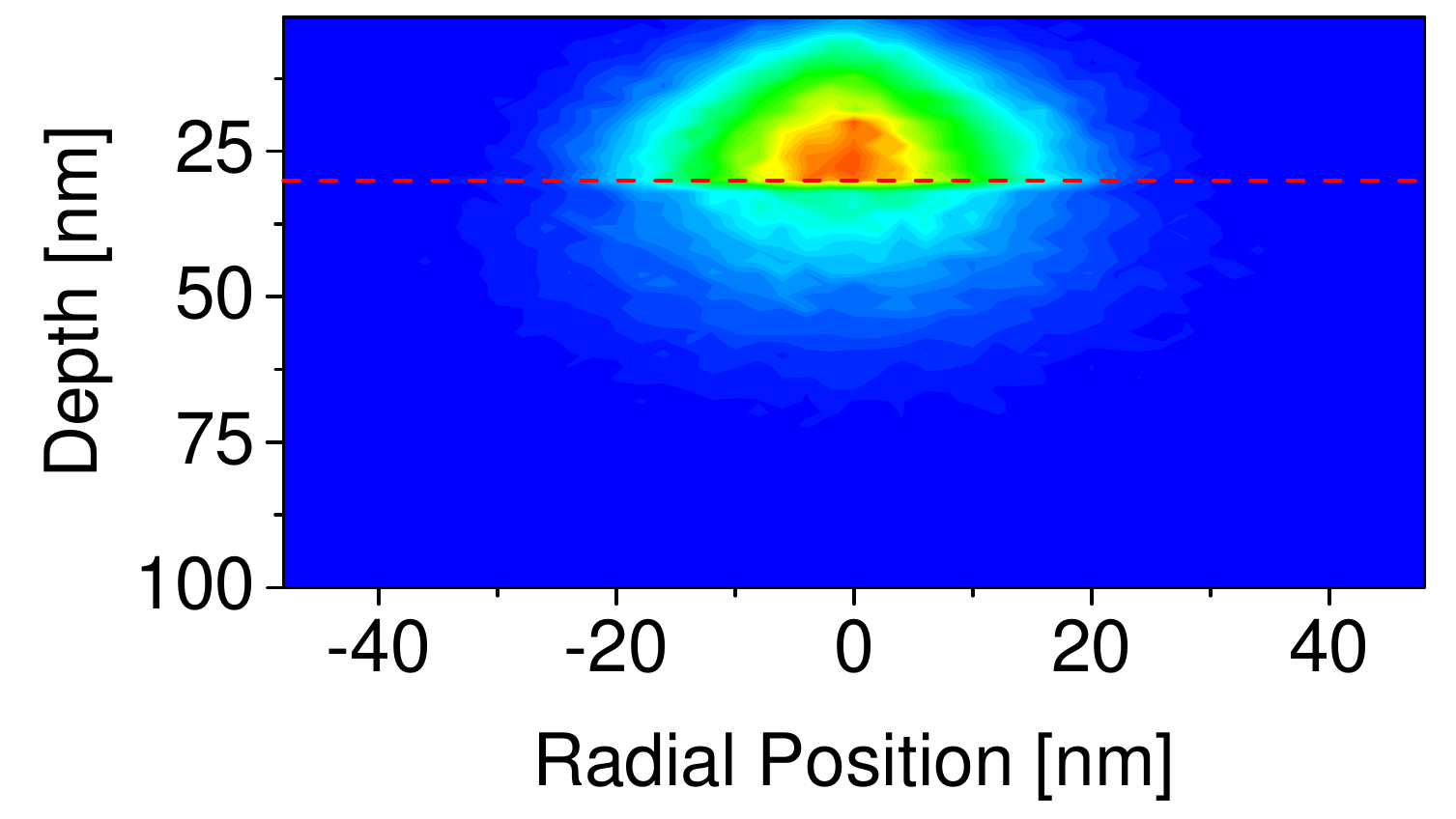}
	}
	\subfloat[]{
		\includegraphics[height=2.98cm]{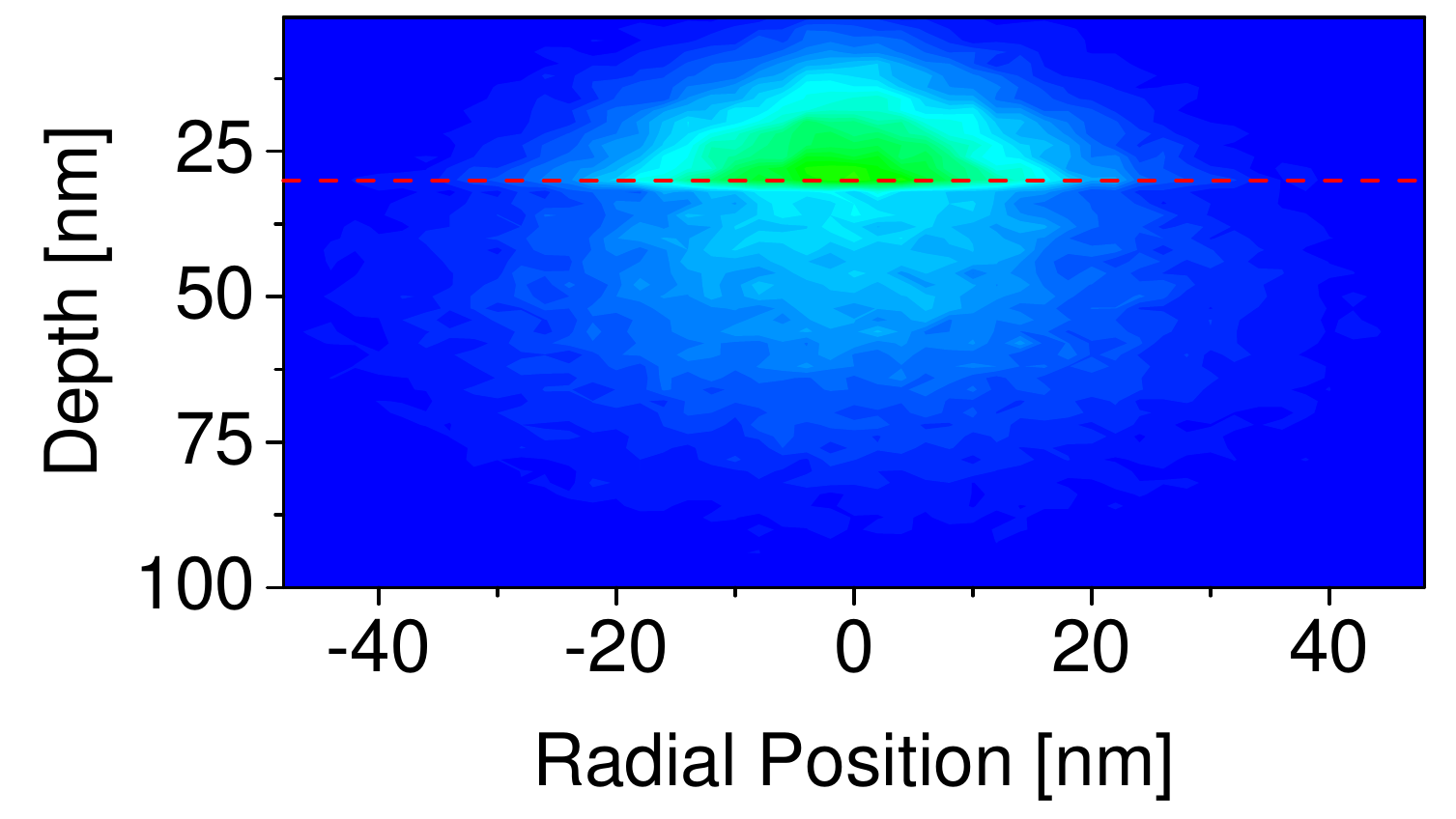}
	}
	\subfloat[]{\raisebox{0mm}{
			\includegraphics[height=2.98cm]{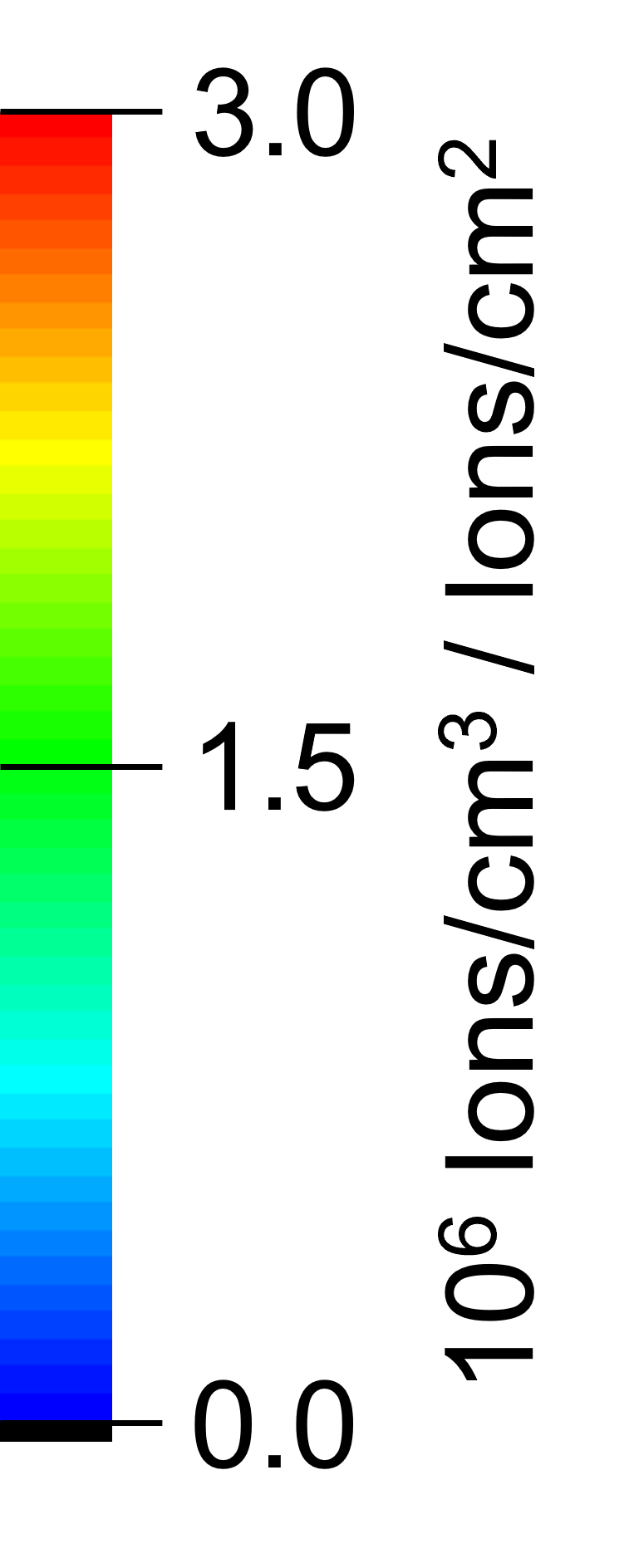}
		}
	}
	\null
	
	\caption{\small{SRIM simulations for 10\textsuperscript{5}\,nitrogen~ions with 10, 15 and 20\,keV (i.e. in our case acceleration energies of 20, 30 and 40\,keV) to illustrate the \textbf{(a-c)} increasing range of vacancies in the silicon lattice and \textbf{(e-g)} final position of nitrogen within the silicon substrate underneath the protective layer of aluminium oxide. The maps within the groups \textbf{(a-c)}\,\&\,\textbf{(e-g)} share the same colour scale, \textbf{(d)} and \textbf{(h)}, respectively, to allow for a better comparison.}}
	\label{fig:Energy-var-2}
\end{figure*}
\noindent
As shown in fig.~\ref{fig:Energy-var-1}~(a)\,\&\,(b), at similar fluence the shape and height of the resulting structures is affected by the energy of the incoming ions.
For acceleration energies from 20 up to 35\,keV we observed a twofold increase of structure height with ion energy.
This originates from the higher ion energy causing more displacements within the silicon, i.e. a larger potential for swelling.
It is illustrated by \textsc{Stopping and Range of Ions in Matter} (SRIM) simulations in fig.~\ref{fig:Energy-var-2}~(a)\Hyphdash*(c) showing the greater amount of displacements reaching into the silicon substrate\textsuperscript{\tiny{\cite{Ziegler2012}}}.
Next to that, we observed distinct features
in the transition region between masked and unmasked areas at higher acceleration energies.
During our experiments we observed the most pronounced features for acceleration energies of 35\,\&\,40\,keV.

Both the appearance of these features and the reduced structure height
for an acceleration energy of 40\,keV can be explained by the formation of Si\textsubscript{x}N\textsubscript{y}\Hyphdash*compounds near the surface of our samples\textsuperscript{\tiny{\cite{Stein1985,Ensinger1998}}}.
This causes an etching of the fabricated structures by phosphoric acid during the removal of the protective aluminium oxide layer, which is likely amplified by the damage caused by the ion irradiation\textsuperscript{\tiny{\cite{Williams2003,Geiss2014,Park2021}}}.
Due to a higher concentration of nitrogen in the centre than in the transition region we expect a faster etch rate in the former.
As the swelling is still pronounced in the transition region due to stress relaxation within the substrate this leads to a reduction of the structure height in the centre below the height in the transition region, i.e. the creation of the observed features.
For an acceleration energy of 40\,keV the spread of ions increases in lateral direction (fig.~\ref{fig:Energy-var-2}~(e)\Hyphdash*(g)), hence the features are more strongly affected by this etching as well.
In combination with the higher damage to the silicon substrate, both measures for structure height are diminished, resulting in a measured height below the values observed for 35\,keV.
One possibility to avoid these effects for all energies would be to forgo the protective layer of aluminium oxide.
However, this can cause contaminations e.g. particles from the chromium etchant to accumulate on the surface\textsuperscript{\tiny{\cite{Bohm2023}}}.

It is important to note that displacements in deeper regions of the substrate likely yield a lower swelling than displacements close to the surface.
This could be an additional factor causing the decrease in height of the structure fabricated with 40\,keV nitrogen ions compared to lower acceleration energies\textsuperscript{\tiny{\cite{Kaufmann2024}}}.
Furthermore, we want to point out that the increase in ion energy leads to an increase in structure width, as expected from the larger lateral spread of displacements in the substrate and final ion position (fig.~\ref{fig:Energy-var-2}).
This can be utilized to tune the duty cycle of fabricated elements, but increases the extent of the transition region and hence the minimal viable structure size slightly.

To minimize the extent of the transition region, the etching and the features related to both, we focused on experiments with an acceleration energy of 20\,keV.
\begin{figure*}[tb]\centering
	\includegraphics[width=0.82\textwidth]{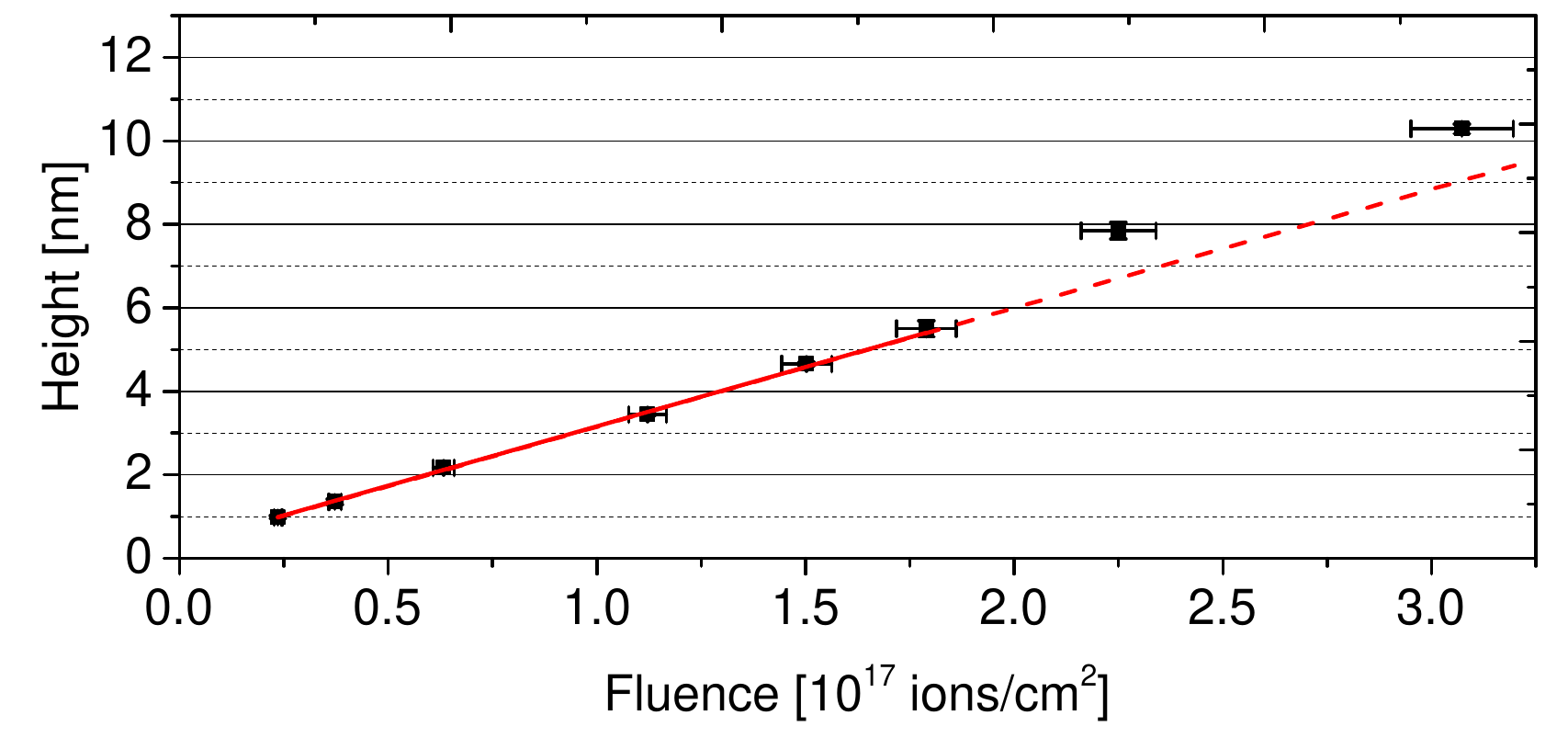}
	\caption{\small{Structure height in dependence of fluence for the irradiation of Si(100) with 20\,keV~N\textsubscript{2}\textsuperscript{+} ions. The solid line is the result of a linear fit of the observed height in dependence of fluence. The dashed line extrapolates this fit as a guide to the eye for comparison with results at higher fluence.}}
	\label{fig:Swelling}
\end{figure*}
\noindent
\begin{figure*}[tb]\centering
	\includegraphics[width=0.82\textwidth]{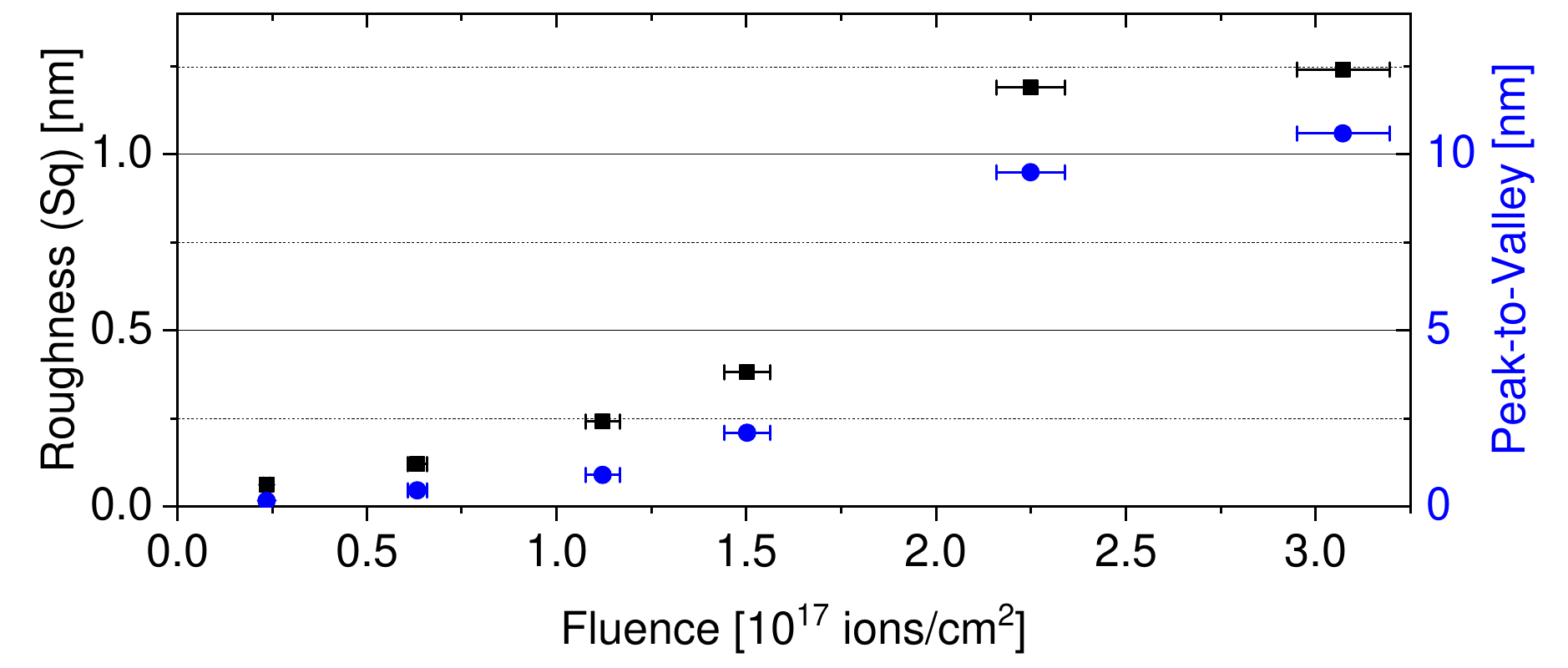}
	\caption{\small{Roughness (Sq) and peak-to-valley values measured in dependence of fluence for the irradiation of
	Si(100) with 20\,keV~N\textsubscript{2}\textsuperscript{+} ions.}}
	\label{fig:Roughness}
\end{figure*}
\noindent
For this energy we measured the linear increase of structure height with fluence shown in fig.~\ref{fig:Swelling}.
With the uncertainty of fluence of the utilized ion source\textsuperscript{\tiny{\cite{Kaufmann2025}}} our technique was suitable to produce gratings with structure heights in the range of 1.00\,$\pm$\,0.05\,nm to 10\,$\pm$\,0.5\,nm.
We want to emphasize that the fluence variations are likely the largest factor in the uncertainty of the fabricated structure height provided above.
This means, improving the accuracy of fluence is a straightforward approach to further improve the fabrication accuracy.

Excluding data for fluence greater than\linebreak 2\,$\cdot$\,10\textsuperscript{17}\,ions\,$\cdot$\,cm\textsuperscript{-2} results in a gradient of\linebreak
2.84\,$\pm$\,0.06\,nm/10\textsuperscript{17}\,ions\,$\cdot$\,cm\textsuperscript{-2} with a y\Hyphdash*intersection at 0.32\,$\pm$\,0.05\,nm.
This indicates the existence of at least two changes in the swelling behaviour up to the maximum investigated fluence:
first, the y\Hyphdash*intersection of larger than 0 means that there is an effect leading to a very strong initial increase before settling for a stable rate.
This is attributed to the increase in substrate temperature during the irradiation, which causes a higher defect recombination rate, i.e. the relative damage yield for the initial ions hitting the sample at close to room temperature is higher\textsuperscript{\tiny{\cite{Pelaz2004}}}.
Second, the jump to increased structure height for fluence greater than 2\,$\cdot$\,10\textsuperscript{17}\,ions\,$\cdot$\,cm\textsuperscript{-2}, which occurs alongside an appearance of features similar to those observed at higher ion energies.
This is remarkable, as for an acceleration energy of 20\,keV atoms in the silicon lattice are displaced by knock on atoms, and nitrogen ions for the most part remain within the protective aluminium oxide layer (see fig.~\ref{fig:Energy-var-2}~(e)).
It is assumed that both effects are linked to the thinning of aluminium oxide via sputtering during the irradiation process, which effectively increases the range of nitrogen and displacements into the silicon substrate, leading to stronger swelling and silicon nitride formation.

With respect to the accelerating increase of roughness (hereafter referring to Sq) with fluence shown in fig.~\ref{fig:Roughness}, it is evident that this effect is of lesser relevance for the applications requiring extremely low roughness considered here, since at high fluence the roughness increased beyond 1\,nm.
The saturation of roughness increase observed for the highest fluence is attributed to the approach of an equilibrium value for the creation and removal of roughness by irradiations with ions.
In contrast to previous similar irradiations utilizing helium ions, which resulted in no relevant increase in roughness\textsuperscript{\tiny{\cite{Kaufmann2025}}}, we now observed a significant and fluence dependent increase in roughness.
This is related to the higher mass of utilized ions, which increases sputter rates and the intermixing and subsequent removal of the silicon substrate with the aluminium oxide layer on top of it.
Depending on the demands of the application regarding roughness this effect limits the accessible structure heights.
Still, for an acceptable roughness of $\leq$\,0.5\,nm, structure heights of greater than 5\,nm are feasible for acceleration energies of 20\,keV.
Additionally, for the utilized ion energies the sputter yield does not increase with increasing ion energy\textsuperscript{\tiny{\cite{Ziegler2012,Wasa2012}}}.
Hence, it is feasible to achieve larger structure heights by increasing the ion energy.

\section{Conclusion}
By irradiation of silicon with varying fluence of N\textsubscript{2}\textsuperscript{+} ions
we were able to demonstrate a process to fabricate
lamellar grating profiles suitable for the EUV and soft X-ray range.
Varying the ion energy yielded an additional parameter affecting shape and height of the resulting gratings.
This expands the possible applications of a previously established technique\textsuperscript{\tiny{\cite{Kaufmann2025}}}.
With respect to the uncertainty of fluence for the utilized ion source, we have shown a reproducible fabrication technique for structure heights in the range of 1.00\,$\pm$\,0.05 to 10\,$\pm$\,0.5\,nm for a pitch of 1\,{\fontfamily{ppl}\selectfont µ}m by irradiation of silicon with 20~N\textsubscript{2}\textsuperscript{+}~ions.
At higher fluence the increase of roughness from 0.06 to 1.25\,nm might hamper the quality of the resulting structures.
However, this can be alleviated by increasing the ion energy to fabricate higher structures at lower fluence.

We have demonstrated the adaptation of a previously established efficient, well controlled and reproducible fabrication technique for gratings applicable in the EUV and soft X-ray range.
Focusing on a pitch of 1\,{\fontfamily{ppl}\selectfont µ}m enabled comparability with previous research and expanded the foundations of the process to other morphologies\textsuperscript{\tiny{\cite{Kaufmann2025}}}.
Future work will investigate the influence of parameters such as pitch and angle of irradiation to enable higher line densities and a wider range of morphologies.
Next to that, an adaptation of the mask to minimize etching of the substrate or possible compounds formed during the irradiation can ease the fabrication further.
Going forward, the presented technique is expected to enable cost-effective access to previously unavailable optics and accelerate the development of novel EUV and soft X-ray devices.

\section*{Acknowledgements}
This work was supported in part by the Fraunhofer-Gesellschaft, the Carl Zeiss Foundation, and the German Federal Ministry of Research, Technology and Space (BMBF grant no. 13N15088).

The sample fabrication within this work was carried out by the microstructure technology team at IAP Jena. The authors would like to thank them for providing the fabrication facilities, carrying out processes and providing support.

\printbibliography

@MastersThesis{Bohm2023,
  author       = {Adrian Bohm},
  school       = {Friedrich-Schiller-University Jena},
  title        = {{Einfluss von Nasschemie und Ionenbestrahlungsprozessen auf die Bildung zeitabh\"angiger nanoskaliger Partikelablagerungen an Siliziumoberfl\"achen}},
  year         = {2023},
  creationdate = {2025-01-29T15:57:33},
}

@Article{Mojarad2015,
  author       = {{Mojarad, N. and Gobrecht, J. and Ekinci, Y.}},
  journal      = {Microelectronic Engineering},
  title        = {{Interference Lithography at EUV and Soft X-Ray Wavelengths: Principles, Methods, and Applications}},
  year         = {2015},
  issn         = {0167-9317},
  note         = {Special Issue on Micro/Nano Lithography with Photons, Electrons \& Ions 2014},
  pages        = {55-63},
  volume       = {143},
  creationdate = {2024-10-13T10:31:55},
  doi          = {https://doi.org/10.1016/j.mee.2015.03.047},
  url          = {https://www.sciencedirect.com/science/article/pii/S0167931715001604},
}

@Article{Levinson2022,
  author       = {Harry J. Levinson},
  journal      = {Japanese Journal of Applied Physics},
  title        = {{High-NA EUV Lithography: Current Status and Outlook for the Future}},
  year         = {2022},
  number       = {SD},
  pages        = {SD0803},
  volume       = {61},
  creationdate = {2024-09-27T13:55:31},
  doi          = {10.35848/1347-4065/ac49fa},
  publisher    = {IOP Publishing},
  url          = {https://dx.doi.org/10.35848/1347-4065/ac49fa},
}

@Article{Fu2023,
  author       = {W E Fu and B C He and W L Wu},
  journal      = {Surface Topography: Metrology and Properties},
  title        = {{The Intensity Enhancement of Transmission Small Angle X-Ray Scattering from Nanostructures with a High Aspect Ratio}},
  year         = {2023},
  number       = {2},
  pages        = {024008},
  volume       = {11},
  creationdate = {2024-10-13T11:26:37},
  doi          = {10.1088/2051-672x/acdcad},
  publisher    = {IOP Publishing},
  url          = {https://dx.doi.org/10.1088/2051-672X/acdcad},
}

@Article{Wu2023,
  author       = {Wen-li Wu and R. Joseph Kline and Ronald L. Jones and Hae-Jeong Lee and Eric K. Lin and Daniel F. Sunday and Chengqing Wang and Tengjiao Hu and Christopher L. Soles},
  journal      = {Journal of Micro/Nanopatterning, Materials, and Metrology},
  title        = {{Review of the Key Milestones in the Development of Critical Dimension Small Angle X-Ray Scattering at National Institute of Standards and Technology}},
  year         = {2023},
  number       = {3},
  pages        = {031206},
  volume       = {22},
  creationdate = {2024-10-13T11:21:57},
  doi          = {10.1117/1.JMM.22.3.031206},
  publisher    = {SPIE},
  url          = {https://doi.org/10.1117/1.JMM.22.3.031206},
}

@InProceedings{Hoenicke2023,
  author       = {Hönicke, Philipp and Kayser, Yves and Soltwisch, Victor and Wählisch, Andre and Wauschkuhn, Nils and Scheerder, Jeroen E. and Fleischmann, Claudia and Bogdanowicz, Janusz and Charley, Anne-Laure and Veloso, Anabela and Loo, Roger and Mertens, Hans and Hikavyy, Andriy and Siefke, Thomas and Andrle, Anna and Gwalt, Grzegorz and Siewert, Frank and Ciesielski, Richard and Beckhoff, Burkhard},
  title        = {{Small Target Compatible Dimensional and Analytical Metrology for Semiconductor Nanostructures Using X-Ray Fluorescence Techniques}},
  year         = {2023},
  pages        = {124961J},
  volume       = {12496},
  creationdate = {2024-10-04T10:40:31},
  doi          = {10.1117/12.2657963},
  journal      = {Proc.SPIE},
  url          = {https://doi.org/10.1117/12.2657963},
}

@Article{Ciesielski2023,
  author       = {Ciesielski, Richard and Lohr, Leonhard M. and Herrero, Analía Fernández and Fischer, Andreas and Grothe, Alexander and Mentzel, Heiko and Scholze, Frank and Soltwisch, Victor},
  journal      = {Review of Scientific Instruments},
  title        = {{A New Sample Chamber for Hybrid Detection of Scattering and Fluorescence, using Synchrotron Radiation in the Soft X-Ray and Extreme Ultraviolet (EUV) Spectral Range}},
  year         = {2023},
  issn         = {0034-6748},
  month        = {01},
  number       = {1},
  pages        = {013904},
  volume       = {94},
  creationdate = {2024-10-04T09:59:55},
  doi          = {10.1063/5.0120146},
  eprint       = {https://pubs.aip.org/aip/rsi/article-pdf/doi/10.1063/5.0120146/16711784/013904\_1\_online.pdf},
  url          = {https://doi.org/10.1063/5.0120146},
}

@Article{Ruben2022,
  author       = {Ruben, Gary and Pinar, Isaac and Brown, Jeremy M. C. and Schaff, Florian and Pollock, James A. and Crossley, Kelly J. and Maksimenko, Anton and Hall, Chris and Hausermann, Daniel and Uesugi, Kentaro and Kitchen, Marcus J.},
  journal      = {IEEE Transactions on Medical Imaging},
  title        = {{Full Field X-Ray Scatter Tomography}},
  year         = {2022},
  number       = {8},
  pages        = {2170-2179},
  volume       = {41},
  creationdate = {2024-10-13T11:17:18},
  doi          = {10.1109/TMI.2022.3157954},
}

@Article{Weinhardt2024,
  author       = {Weinhardt, Lothar and Wansorra, Constantin and Steininger, Ralph and Spangenberg, Thomas and Hauschild, Dirk and Heske, Clemens},
  journal      = {Journal of Synchrotron Radiation},
  title        = {{High-Transmission Spectrometer for Rapid Resonant Inelastic Soft X-Ray Scattering (rRIXS) Maps}},
  year         = {2024},
  number       = {6},
  pages        = {1481--1488},
  volume       = {31},
  creationdate = {2025-01-30T10:08:26},
  doi          = {10.1107/S160057752400804X},
  url          = {https://doi.org/10.1107/S160057752400804X},
}

@Article{Shatokhin2018,
  author       = {A. N. Shatokhin and A. O. Kolesnikov and P. V. Sasorov and E. A. Vishnyakov and E. N. Ragozin},
  journal      = {Opt. Express},
  title        = {{High-Resolution Stigmatic Spectrograph for a Wavelength Range of 12.5-30 nm}},
  year         = {2018},
  number       = {15},
  pages        = {19009--19019},
  volume       = {26},
  creationdate = {2025-01-30T10:13:43},
  doi          = {10.1364/OE.26.019009},
  publisher    = {Optica Publishing Group},
  url          = {https://opg.optica.org/oe/abstract.cfm?URI=oe-26-15-19009},
}

@Article{Jiang2013,
  author       = {Jiang, Y H and Senftleben, A and Kurka, M and Rudenko, A and Foucar, L and Herrwerth, O and Kling, M F and Lezius, M and Tilborg, J V and Belkacem, A and Ueda, K and Rolles, D and Treusch, R and Zhang, Y Z and Liu, Y F and Schröter, C D and Ullrich, J and Moshammer, R},
  journal      = {Journal of Physics B: Atomic, Molecular and Optical Physics},
  title        = {{Ultrafast Dynamics in Acetylene Clocked in a Femtosecond XUV Stopwatch}},
  year         = {2013},
  number       = {16},
  pages        = {164027},
  volume       = {46},
  creationdate = {2025-04-17T12:33:15},
  doi          = {10.1088/0953-4075/46/16/164027},
  publisher    = {IOP Publishing},
  url          = {https://dx.doi.org/10.1088/0953-4075/46/16/164027},
}

@Article{Wang2023,
  author       = {Wang, Enliang and Kling, Nora G. and LaForge, Aaron C. and Obaid, Razib and Pathak, Shashank and Bhattacharyya, Surjendu and Meister, Severin and Trost, Florian and Lindenblatt, Hannes and Schoch, Patrizia and Kübel, Matthias and Pfeifer, Thomas and Rudenko, Artem and Díaz-Tendero, Sergio and Martín, Fernando and Moshammer, Robert and Rolles, Daniel and Berrah, Nora},
  journal      = {J. Phys. Chem. Lett.},
  title        = {{Ultrafast Roaming Mechanisms in Ethanol Probed by Intense Extreme Ultraviolet Free-Electron Laser Radiation: Electron Transfer versus Proton Transfer}},
  year         = {2023},
  month        = may,
  number       = {18},
  pages        = {4372--4380},
  volume       = {14},
  comment      = {doi: 10.1021/acs.jpclett.2c03764},
  creationdate = {2025-04-17T11:47:29},
  doi          = {10.1021/acs.jpclett.2c03764},
  publisher    = {American Chemical Society},
  url          = {https://doi.org/10.1021/acs.jpclett.2c03764},
}

@Article{Aidukas2024,
  author       = {Aidukas, Tomas and Phillips, Nicholas W. and Diaz, Ana and Poghosyan, Emiliya and Müller, Elisabeth and Levi, A. F. J. and Aeppli, Gabriel and Guizar-Sicairos, Manuel and Holler, Mirko},
  journal      = {Nature},
  title        = {{High-Performance 4-nm-Resolution X-Ray Tomography using Burst Ptychography}},
  year         = {2024},
  issn         = {1476-4687},
  number       = {8023},
  pages        = {81--88},
  volume       = {632},
  creationdate = {2024-10-13T11:04:06},
  doi          = {10.1038/s41586-024-07615-6},
  refid        = {Aidukas2024},
  url          = {https://doi.org/10.1038/s41586-024-07615-6},
}

@Article{Battistelli2024,
  author       = {Riccardo Battistelli and Daniel Metternich and Michael Schneider and Lisa-Marie Kern and Kai Litzius and Josefin Fuchs and Christopher Klose and Kathinka Gerlinger and Kai Bagschik and Christian M. G{\"u}nther and Dieter Engel and Claus Ropers and Stefan Eisebitt and Bastian Pfau and Felix B{\"u}ttner and Sergey Zayko},
  journal      = {Optica},
  title        = {{Coherent X-Ray Magnetic Imaging with 5 nm Resolution}},
  year         = {2024},
  number       = {2},
  pages        = {234--237},
  volume       = {11},
  creationdate = {2024-10-04T11:33:52},
  doi          = {10.1364/OPTICA.505999},
  publisher    = {Optica Publishing Group},
  url          = {https://opg.optica.org/optica/abstract.cfm?URI=optica-11-2-234},
}

@Article{Grote2022,
  author       = {{Grote, Lukas et al.}},
  journal      = {Nature Communications},
  title        = {{Imaging Cu{\textsubscript{2}}O Nanocube Hollowing in Solution by Quantitative In Situ X-Ray Ptychography}},
  year         = {2022},
  issn         = {2041-1723},
  number       = {1},
  pages        = {4971},
  volume       = {13},
  creationdate = {2024-10-13T11:12:05},
  doi          = {10.1038/s41467-022-32373-2},
  refid        = {Grote2022},
  url          = {https://doi.org/10.1038/s41467-022-32373-2},
}

@Article{Gutberlet2023,
  author       = {Tobias Gutberlet and Hung-Tzu Chang and Sergey Zayko and Murat Sivis and Claus Ropers},
  journal      = {Opt. Express},
  title        = {{High-Sensitivity Extreme-Ultraviolet Transient Absorption Spectroscopy Enabled by Machine Learning}},
  year         = {2023},
  number       = {24},
  pages        = {39757--39764},
  volume       = {31},
  creationdate = {2024-10-04T11:29:15},
  doi          = {10.1364/OE.495821},
  publisher    = {Optica Publishing Group},
  url          = {https://opg.optica.org/oe/abstract.cfm?URI=oe-31-24-39757},
}

@Article{Li2024,
  author       = {{Li, Jie et al.}},
  journal      = {Light: Science {\&} Applications},
  title        = {{Highly Efficient and Aberration-Free off-Plane Grating Spectrometer and Monochromator for EUV-Soft X-Ray Applications}},
  year         = {2024},
  issn         = {2047-7538},
  number       = {1},
  pages        = {12},
  volume       = {13},
  creationdate = {2024-10-04T11:45:24},
  doi          = {10.1038/s41377-023-01342-9},
  refid        = {Li2024},
  url          = {https://doi.org/10.1038/s41377-023-01342-9},
}

@Article{Poletto2018,
  author    = {Luca Poletto and Fabio Frassetto},
  journal   = {Appl. Opt.},
  title     = {{Cost-Effective Plane-Grating Monochromator Design for Extreme-Ultraviolet Application}},
  year      = {2018},
  number    = {5},
  pages     = {1202--1211},
  volume    = {57},
  doi       = {10.1364/AO.57.001202},
  publisher = {OSA},
  url       = {http://www.osapublishing.org/ao/abstract.cfm?URI=ao-57-5-1202},
}

@Article{Wu2010,
  author       = {Wu, Banqiu and Kumar, Ajay and Pamarthy, Sharma},
  journal      = {Journal of Applied Physics},
  title        = {{High Aspect Ratio Silicon Etch: A Review}},
  year         = {2010},
  issn         = {0021-8979},
  month        = {09},
  number       = {5},
  pages        = {051101},
  volume       = {108},
  creationdate = {2024-09-30T10:57:37},
  doi          = {10.1063/1.3474652},
  eprint       = {https://pubs.aip.org/aip/jap/article-pdf/doi/10.1063/1.3474652/14067262/051101\_1\_online.pdf},
  url          = {https://doi.org/10.1063/1.3474652},
}

@Article{Dowling2016,
  author       = {Dowling, Karen and Ransom, Elliot and Senesky, Debbie},
  journal      = {Journal of Microelectromechanical Systems},
  title        = {{Profile Evolution of High Aspect Ratio Silicon Carbide Trenches by Inductive Coupled Plasma Etching}},
  year         = {2016},
  month        = {11},
  pages        = {1-8},
  volume       = {PP},
  creationdate = {2024-09-30T11:00:57},
  doi          = {10.1109/JMEMS.2016.2621131},
}

@Article{RackaSzmidt2022,
  author         = {Racka-Szmidt, Katarzyna and Stonio, Bartłomiej and Zelazko, Jarosław and Filipiak, Maciej and Sochacki, Mariusz},
  journal        = {Materials},
  title          = {{A Review: Inductively Coupled Plasma Reactive Ion Etching of Silicon Carbide}},
  year           = {2022},
  issn           = {1996-1944},
  number         = {1},
  volume         = {15},
  article-number = {123},
  creationdate   = {2024-09-30T11:03:38},
  doi            = {10.3390/ma15010123},
  pubmedid       = {35009277},
  url            = {https://www.mdpi.com/1996-1944/15/1/123},
}

@Article{Chen2021,
  author    = {Yuheng Chen and Maoqi Cai and Haofeng Zang and Huoyao Chen and Stefanie Kroker and Yonghua Lu and Ying Liu and Frank Frost and Yilin Hong},
  journal   = {Appl. Opt.},
  title     = {{Optical Anisotropy of Self-Organized Gold Quasi-Blazed Nanostructures Based on a Broad Ion Beam}},
  year      = {2021},
  number    = {3},
  pages     = {505--512},
  volume    = {60},
  doi       = {10.1364/AO.412631},
  publisher = {Optica Publishing Group},
  url       = {https://opg.optica.org/ao/abstract.cfm?URI=ao-60-3-505},
}

@InProceedings{Miles2022,
  author       = {Drew M. Miles and Randall L. McEntaffer and Fabien Gris{\'e}},
  booktitle    = {Space Telescopes and Instrumentation 2022: Ultraviolet to Gamma Ray},
  title        = {{Blazed Reflection Gratings with Electron-Beam Lithography and Ion-Beam Etching}},
  year         = {2022},
  editor       = {Jan-Willem A. den Herder and Shouleh Nikzad and Kazuhiro Nakazawa},
  organization = {International Society for Optics and Photonics},
  pages        = {1218153},
  publisher    = {SPIE},
  volume       = {12181},
  doi          = {10.1117/12.2637880},
  url          = {https://doi.org/10.1117/12.2637880},
}

@InCollection{Wasa2012,
  author       = {Kiyotaka Wasa},
  booktitle    = {{Handbook of Sputtering Technology (Second Edition)}},
  publisher    = {William Andrew Publishing},
  title        = {2 - Sputtering Phenomena},
  year         = {2012},
  address      = {Oxford},
  edition      = {Second Edition},
  editor       = {Kiyotaka Wasa and Isaku Kanno and Hidetoshi Kotera},
  isbn         = {978-1-4377-3483-6},
  pages        = {41-75},
  creationdate = {2025-01-29T13:19:53},
  doi          = {https://doi.org/10.1016/B978-1-4377-3483-6.00002-4},
  url          = {https://www.sciencedirect.com/science/article/pii/B9781437734836000024},
}

@Article{Geiss2014,
  author       = {Geiss, Reinhard and Brandt, Juliane and Hartung, Holger and Tünnermann, Andreas and Pertsch, Thomas and Kley, Ernst-Bernhard and Schrempel, Frank},
  journal      = {Journal of Vacuum Science \& Technology B},
  title        = {{Photonic Microstructures in Lithium Niobate by Potassium Hydroxide-Assisted ion Beam-Enhanced Etching}},
  year         = {2014},
  issn         = {2166-2746},
  month        = {11},
  number       = {1},
  pages        = {010601},
  volume       = {33},
  creationdate = {2025-01-31T08:12:41},
  doi          = {10.1116/1.4902087},
  eprint       = {https://pubs.aip.org/avs/jvb/article-pdf/doi/10.1116/1.4902087/14695368/010601\_1\_online.pdf},
  url          = {https://doi.org/10.1116/1.4902087},
}

@Article{Williams2003,
  author  = {Williams, K.R. and Gupta, K. and Wasilik, M.},
  journal = {Journal of Microelectromechanical Systems},
  title   = {{Etch Rates for Micromachining Processing-Part II}},
  year    = {2003},
  number  = {6},
  pages   = {761-778},
  volume  = {12},
  doi     = {10.1109/JMEMS.2003.820936},
}

@Article{Park2021,
  author       = {Soyong Park and Hyunwook Jung and Kyung-Ah Min and Junyeop Kim and Byungchan Han},
  journal      = {Applied Surface Science},
  title        = {{Unraveling the Selective Etching Mechanism of Silicon Nitride over Silicon Dioxide by Phosphoric Acid: First-Principles Study}},
  year         = {2021},
  issn         = {0169-4332},
  pages        = {149376},
  volume       = {551},
  creationdate = {2025-01-31T08:18:39},
  doi          = {https://doi.org/10.1016/j.apsusc.2021.149376},
  url          = {https://www.sciencedirect.com/science/article/pii/S0169433221004529},
}

@Article{Stein1985,
  author  = {Stein, Herman J.},
  journal = {MRS Proceedings},
  title   = {{Nitrogen in Crystalline Si}},
  year    = {1985},
  pages   = {523},
  volume  = {59},
  doi     = {10.1557/PROC-59-523},
}

@Article{Ensinger1998,
  author       = {Ensinger, W. and Volz, K. and Schrag, G. and Stritzker, B. and Rauschenbach, B.},
  journal      = {Applied Physics Letters},
  title        = {{Formation of Silicon Nitride Layers by Nitrogen Ion Irradiation of Silicon biased to a High Voltage in an Electron Cyclotron Resonance Microwave Plasma}},
  year         = {1998},
  issn         = {0003-6951},
  month        = {03},
  number       = {10},
  pages        = {1164-1166},
  volume       = {72},
  creationdate = {2025-01-31T08:21:14},
  doi          = {10.1063/1.121001},
  eprint       = {https://pubs.aip.org/aip/apl/article-pdf/72/10/1164/18532514/1164\_1\_online.pdf},
  url          = {https://doi.org/10.1063/1.121001},
}

@Book{Ziegler2012,
  author    = {Ziegler, J. F. and Biersack J. P. and Ziegler, M. D.},
  publisher = {Chester, Md, SRIM Co.},
  title     = {{SRIM: the Stopping and Range of Ions in Matter}},
  year      = {2012},
}

@Article{Kaufmann2025,
  author       = {Kaufmann, J and Ciesielski, R and Freiberg, K and Walther, M and Fernández Herrero, A and Lippmann, S and Soltwisch, V and Siefke, T and Zeitner, U},
  journal      = {Nanotechnology},
  title        = {{Fabrication of Shallow EUV Gratings on Silicon by Irradiation with Helium Ions}},
  year         = {2025},
  number       = {18},
  pages        = {185301},
  volume       = {36},
  creationdate = {2025-04-14T08:50:58},
  doi          = {10.1088/1361-6528/adc4ec},
  publisher    = {IOP Publishing},
  url          = {https://dx.doi.org/10.1088/1361-6528/adc4ec},
}

@InProceedings{Kaufmann2025a,
  author       = {Johannes Kaufmann and Richard Ciesielski and Katharina Freiberg and Markus Walther and Anal{\'i}a Fern{\'a}ndez Herrero and Stephanie Lippmann and Victor Soltwisch and Thomas Siefke and Uwe Zeitner},
  booktitle    = {EUV and X-ray Optics: Synergy between Laboratory and Space IX},
  title        = {{Ultra Shallow Silicon EUV Gratings Fabricated via Ion Irradiation}},
  year         = {2025},
  editor       = {Ren{\'e} Hudec and Ladislav Pina},
  organization = {International Society for Optics and Photonics},
  pages        = {135310L},
  publisher    = {SPIE},
  volume       = {13531},
  doi          = {10.1117/12.3055797},
  url          = {https://doi.org/10.1117/12.3055797},
}

@InProceedings{Kaufmann2025b,
  author       = {Johannes Kaufmann and Richard Ciesielski and Katharina Freiberg and Markus Walther and Anal{\'i}a Fern{\'a}ndez Herrero and Stephanie Lippmann and Victor Soltwisch and Thomas Siefke and Uwe Zeitner},
  booktitle    = {Advances in X-Ray/EUV Sources, Optics, and Components XX},
  title        = {{Fabrication of Ultra-Shallow EUV Gratings in Silicon via Ion Irradiation}},
  year         = {2025},
  editor       = {Ali M. Khounsary and Hidekazu Mimura},
  organization = {International Society for Optics and Photonics},
  pages        = {136200I},
  publisher    = {SPIE},
  volume       = {13620},
  doi          = {10.1117/12.3061518},
  url          = {https://doi.org/10.1117/12.3061518},
}

@Article{Pelaz2004,
  author  = {Pelaz,Lourdes and Marqués,Luis A. and Barbolla,Juan},
  journal = {Journal of Applied Physics},
  title   = {{Ion-Beam-Induced Amorphization and Recrystallization in Silicon}},
  year    = {2004},
  number  = {11},
  pages   = {5947-5976},
  volume  = {96},
  doi     = {10.1063/1.1808484},
  eprint  = {https://doi.org/10.1063/1.1808484},
  url     = {https://doi.org/10.1063/1.1808484},
}

@Article{Jacob1998,
  author       = {W Jacob},
  journal      = {Thin Solid Films},
  title        = {{Surface Reactions during Growth and Erosion of Hydrocarbon Films}},
  year         = {1998},
  issn         = {0040-6090},
  number       = {1},
  pages        = {1-42},
  volume       = {326},
  creationdate = {2025-01-29T09:14:54},
  doi          = {https://doi.org/10.1016/S0040-6090(98)00497-0},
  url          = {https://www.sciencedirect.com/science/article/pii/S0040609098004970},
}

@Article{Garcia2022,
  author   = {García, G. and Martin, M. and Ynsa, M. D. and Torres-Costa, V. and Crespillo, M. L. and Tardío, M. and Olivares, J. and Bosia, F. and Peña-Rodríguez, O. and Nicolas, J. and Tallarida, M.},
  journal  = {The European Physical Journal Plus},
  title    = {{Process Design for the Manufacturing of Soft X-Ray Gratings in Single-Crystal Diamond by High-Energy Heavy-Ion Irradiation}},
  year     = {2022},
  issn     = {2190-5444},
  number   = {10},
  pages    = {1157},
  volume   = {137},
  doi      = {10.1140/epjp/s13360-022-03358-3},
  refid    = {García2022},
  url      = {https://doi.org/10.1140/epjp/s13360-022-03358-3},
}

@Article{Yang2017a,
  author       = {Xiaowei Yang and Igor V. Kozhevnikov and Qiushi Huang and Hongchang Wang and Matthew Hand and Kawal Sawhney and Zhanshan Wang},
  journal      = {Opt. Express},
  title        = {{Analytic Theory of Alternate Multilayer Gratings Operating in Single-Order Regime}},
  year         = {2017},
  number       = {14},
  pages        = {15987--16001},
  volume       = {25},
  creationdate = {2024-12-19T10:43:09},
  doi          = {10.1364/OE.25.015987},
  publisher    = {Optica Publishing Group},
  url          = {https://opg.optica.org/oe/abstract.cfm?URI=oe-25-14-15987},
}

@Article{Koike2023,
  author       = {Koike, M. and Hatano, T. and Pirozhkov, A. S. and Ueno, Y. and Terauchi, M.},
  journal      = {Review of Scientific Instruments},
  title        = {{Design of Soft X-Ray Laminar-Type Gratings Coated with Supermirror-Type Multilayer to Enhance Diffraction Efficiency in a Region of 2–4 keV}},
  year         = {2023},
  issn         = {0034-6748},
  month        = {04},
  number       = {4},
  pages        = {045109},
  volume       = {94},
  creationdate = {2024-12-18T15:41:13},
  doi          = {10.1063/5.0148908},
  eprint       = {https://pubs.aip.org/aip/rsi/article-pdf/doi/10.1063/5.0148908/16833483/045109\_1\_5.0148908.pdf},
  url          = {https://doi.org/10.1063/5.0148908},
}

@Article{Koike2024a,
  author       = {Koike, M. and Hatano, T. and Pirozhkov, A. S. and Oue, Y. and Murano, T. and Kakio, T. and Koshiya, S. and Kondo, K. and Terauchi, M.},
  journal      = {Review of Scientific Instruments},
  title        = {{Soft X-Ray High Diffraction Efficiency and Spectral Flux Laminar-Rype W/B4C Multilayer Diffraction Grating for 300–1000 eV}},
  year         = {2024},
  issn         = {0034-6748},
  month        = {07},
  number       = {7},
  pages        = {073104},
  volume       = {95},
  creationdate = {2024-12-19T09:07:17},
  doi          = {10.1063/5.0210722},
  eprint       = {https://pubs.aip.org/aip/rsi/article-pdf/doi/10.1063/5.0210722/20069792/073104\_1\_5.0210722.pdf},
  url          = {https://doi.org/10.1063/5.0210722},
}

@Article{Momota2019,
  author   = {S. Momota and N. Sato and K. Honda},
  journal  = {Vacuum},
  title    = {{Fabrication of Multi-Step Swelling Structures on 6H–SiC by Using Highly-Charged Ar Beams}},
  year     = {2019},
  issn     = {0042-207X},
  pages    = {108963},
  volume   = {170},
  doi      = {https://doi.org/10.1016/j.vacuum.2019.108963},
  url      = {https://www.sciencedirect.com/science/article/pii/S0042207X19306116},
}

@Article{Zhou2019,
  author  = {Yuying Zhou and Shimin Li and Ying Wang and Qing Huang and Wei Zhang and Yao Yao and Jiaming Hao and Yan Sun and Ming Tang and Bin Li and Yi Zhang and Jun Hu and Long Yan},
  journal = {Carbon},
  title   = {{One-Step Ion Beam Irradiation Manufacture of 3D Micro/Nanopatterned Structures in SiC with Tunable Work Functions}},
  year    = {2019},
  issn    = {0008-6223},
  pages   = {387-393},
  volume  = {148},
  doi     = {https://doi.org/10.1016/j.carbon.2019.04.011},
  url     = {https://www.sciencedirect.com/science/article/pii/S0008622319303288},
}

@Article{Jensen2008,
  author   = {J. Jensen and M. Skupiński and K. Hjort and R. Sanz},
  journal  = {Nuclear Instruments and Methods in Physics Research Section B: Beam Interactions with Materials and Atoms},
  title    = {{Heavy Ion Beam-Based Nano- and Micro-Structuring of TiO\textsubscript{2} Single Crystals using Self-Assembled Masks}},
  year     = {2008},
  issn     = {0168-583X},
  note     = {Radiation Effects in Insulators},
  number   = {12},
  pages    = {3113-3119},
  volume   = {266},
  doi      = {https://doi.org/10.1016/j.nimb.2008.03.220},
  url      = {https://www.sciencedirect.com/science/article/pii/S0168583X08004576},
}


\end{document}